\documentstyle[preprint,aps]{revtex}
\begin{document}

\draft

\title{Conditions for Isoscaling in Nuclear Reactions}
\author{M.~B. Tsang$^{1}$, W.A. Friedman$^{2}$, C.K. Gelbke$^{1}$,
  W.G. Lynch$^{1}$, G. Verde$^{1}$, H.S. Xu$^{1}$}
\author{\small $^{1}$ National Superconducting Cyclotron Laboratory
  and Department of Physics and Astronomy, Michigan State University, East Lansing, MI 48824}
\author{\small $^{2}$ Department of Physics, University of Wisconsin,
  Madison, WI 53706}

\date{\today}

\maketitle

\begin{abstract}
Isoscaling, where ratios of isotopes emitted from two reactions exhibit an
exponential dependence on the neutron and proton number of the
isotope, has been observed over a variety of reactions including evaporation, strongly
damped binary collision, and multifragmentation. The conditions for
isoscaling to occur as well as the conditions when isoscaling fails are
investigated.
\end{abstract}

With the availability of rare isotope beams as well as detection systems
that can resolve the masses and charges of the detected particles, isotope
yields become an important observable for studying nuclear collisions of
heavy ions [1,2]. This additional freedom on isospin asymmetry allows one to
study the properties of bulk nuclear matter that are affected by the nucleon
composition of the nuclei such as the isospin dependence of the liquid gas
phase transition of nuclear matter [3,4] and the asymmetry term [5-8] in the
nuclear equation of state. To minimize undesirable complications stemming
from the sequential decays of primary unstable fragments, it has been
proposed that isospin effects can best be studied by comparing the same
observables in two similar reactions that differ mainly in isospin asymmetry
[4,6,8]. If two reactions, 1 and 2, have the same temperature but different
isospin asymmetry, for example, the ratios of a specific isotope yield with
neutron and proton number $N$ and $Z$, obtained from system 2 and system 1
have been observed to exhibit isoscaling i.e. exponential dependence of the
form [4,5]:

\begin{equation}
R_{21}(N,Z)=Y_{2}(N,Z)/Y_{1}(N,Z)=C\cdot \mathrm{exp}(N\cdot \alpha +Z\cdot
\beta )  
\label{eq:alphabeta}
\end{equation}

where $\alpha $ and $\beta $ are the scaling parameters and $C$ is an
overall normalization constant. We adopt the convention that the neutron and
proton composition of reaction 2 to be more neutron-rich than that of
reaction 1.

Figure 1 illustrates the isoscaling property observed with the fragments
produced in the central collisions of $^{124}Sn+^{124}Sn$ and $%
^{112}Sn+^{112}Sn$ reactions [4]. The $N$ and $Z$ dependence of Eq. 
\ref{eq:alphabeta} becomes most apparent if $R_{21}(N,Z)$
is plotted versus $N$ or $Z$ on a semi-log plot as shown in the
left panels. Isotopes of the same elements are plotted with the same
symbols. Odd-Z nuclei are represented by open
symbols while the even-Z nuclei are represented by closed symbols. 
In the upper left panel, the isotopes for each element Z appear to 
lie on one line and the
resulting slopes would then be $\alpha $. The dashed (for odd-Z
elements) and solid (even-Z elements) lines are best fits to the data points
with one common $\alpha $ values for all the elements. In this case, $\alpha
=0.361$. Similarly, plotting $R_{21}(N,Z)$ against Z for all isotones would
provide a common slope $\beta $ for each $N$. This is demonstrated in the
lower left panel of Figure 1. The best fits (dashed and solid lines) yield a
value of $\beta =-0.417$. The positive (negative) slopes of the lines in the
top (bottom) panel arise from the fact that more neutron-rich (proton-rich)
nuclei are produced in the more neutron-rich (proton-rich) system which
represent the values in the numerators (denominators) of the ratios in Eq. 1.

Alternatively, the data in the left panels can be displayed compactly as a
function of one variable, either $N$ or $Z$, by removing the dependence of
the other variable using the scaled isotope or isotone functions [6]:

\begin{equation}
S(N)=R_{21}(N,Z)\times \mathrm{exp}(-\beta Z).  
\label{eq:SN}
\end{equation}
\begin{equation}
S(Z)=R_{21}(N,Z)\times \mathrm{exp}(-\alpha N).  
\label{eq:SZ}
\end{equation}

For the best fit value of $\beta(=-0.417)$, $S(N)$ 
for all elements lies on a single
line on a semi-log plot as a function of $N$ as shown in the upper right
panel of Figure 1. Alternatively, $S(Z)$ of all isotones
lies on a single line on a semi-log plot as a function of $Z$ for the best
fit value of $\alpha (=0.361)$ as shown in the lower right panel. Both solid
lines shown in the right panels have the same exponential dependence on $N$
and $Z$ as the corresponding lines in the left panels. The agreement between
the data and the lines is excellent, verifying the scaling relation of Eq.
1-3. In general, the fit for $S(N)$\ is better than $S(Z);$ this may reflect
the influence of Coulomb interaction which may not be well approximated by
Eq. 1 [6,8]. On the other hand, $S(N)$ is affected mainly by the differences
in neutron chemical potentials or the neutron separation energies. These
latter factors may be governed by the differences in the symmetry energies
in the two systems [8].

In a recent survey of heavy ion induced reactions, isoscaling appears to be
manifested in a variety of nuclear reactions including deeply inelastic
collisions, evaporation and multifragmentation over a wide range of incident
energies [6]. In this article, we will perform a comprehensive
exploration of many reactions and examine conditions under which isoscaling
is observed and others where it is not. We will also demonstrate how
isoscaling can be restored even when two systems have different temperatures.

{\sc I. Deeply Inelastic Collisions}

In the 1970's, the Deeply Inelastic Collision (DIC) phenomenon was
discovered when heavy ions were used to bombard heavy targets in an effort
to create superheavy elements [9, 10]. Products from the DIC exhibit
characteristics which can partly be attributed to compound nuclei decay and
partly to multinucleon transfer reactions depending on the incident energy
and detection angles. In ref. [6] the isotope yield ratios of $^{16}O$ + $%
^{232}Th$ and $^{16}O$ + $^{197}Au$ reactions at incident energy of 137 MeV
and $\theta =40^{o}$ have been found to exhibit isoscaling behavior. From
the literature, we have selected four additional systems to illustrate the
compliance or noncompliance of the scaling behavior in DIC . Each pair of
the chosen reactions use the same projectile at the same incident energy and
detect the isotopes at the same laboratory angles. The main differences are
the targets. Figure 2 shows the relative isotope ratios, $R_{21}(N,Z)$ for
the four systems: a.)$^{16}O$ + $^{232}Th$ [9] and $^{16}O$ + $^{197}Pb$
[10] at incident energy of 137 MeV and $\theta =40^{o}$ (upper left panel),
b.) $^{14}N$ + $^{100}Mo$ and $^{14}N$ + $^{92}Mo$ at 97 MeV and $\theta
=25^{o}$ (upper right panel) [9], c.) $^{22}Ne$ + $^{232}Th$ and $^{22}Ne$ +
$^{94}Zr$ at 173 MeV and $\theta =12^{o}$ [9] (lower left panel), and d.) $%
^{16}O$ + $^{232}Th$ and $^{16}O$ + $^{197}Au$ at 315 MeV and $\theta =40^{o}
$ (lower right panel) [10]. Isotopes of the same elements are plotted with
the same symbols using the same convention as Figure 1, open circles, closed
circles, open square, closed squares, and open diamonds for Z=3, 4, 5, 6,
and 7, respectively. The solid and dotted lines connect isotopes of the same
elements (solid lines for even Z element and dashed line for odd Z
elements). Scaling similar to Equation 1 is observed for the isotope ratios
plotted in the upper panels. When the product nuclei are detected at very
forward angles, such as the isotope ratios from the $^{22}Ne$ induced
reactions shown in the lower left panel, scaling is not observed. When the
incident energy is raised to 20 MeV per nucleon, target dependence is much
weaker than at lower energies [10] and production of isotopes at forward
angles is consistent with fragmentation of the projectile and shows no
target dependence. In that case, $\alpha =0,\beta =0$ and $%
R_{21}(N,Z)\approx 1$ is observed (lower right panel).

Figure 2 summarizes that isotopic scaling is reasonably well respected at
low incident energies ($E/A<10MeV$) and at angles backward of the grazing
angles but poorly respected at forward angles. The situation at higher
energies is not clear. Isoscaling may have been observed with very small
values of $\alpha $ and $\beta .$ The positive observation of isoscaling can
be understood as follows: Backward of the grazing angles, it is often
assumed that equilibrium is established between the orbiting projectile and
target. In such cases, the isotopic yields follow the ''$Q_{gg}$%
-systematics''[9,10], in which the primary isotope yield of the
projectile-like fragment depends mainly on the $Q$-value of the mass
transfer and can be approximated by

\begin{equation}
Y(N,Z)\propto \mathrm{exp}((M_{P}+M_{T}-M_{P}^{\prime }-M_{T}^{\prime })/T)
\label{eq:Qgg}
\end{equation}

\noindent where $M_{P}$ and $M_{T}$ are the initial projectile and target
masses, and $M_{P}^{\prime }$ and $M_{T}^{\prime }$ are the final masses of
the projectile- and target-like fragment. Here, $T$ has a natural
interpretation as the temperature, but is not always assumed to be so.
Applying charge and mass conservation, and expressing explicitly only the
terms that depend on $N$ and $Z$, one can write $R_{21}(N,Z)$ as

\begin{equation}
R_{21}(N, Z)\propto \exp [(BE(N_{2}-N,Z_{2}-Z)-BE(N_{1}-N,Z_{1}-Z))/T],
\label{eq:R21_Qgg}
\end{equation}

\noindent where $Z_{i}$ and $N_{i}$ are the total proton and neutron number
of reaction $i$. $BE$ is the binding energy of a nucleus. Expanding the
binding energies in Taylor series, one obtains an expression of the form

\begin{equation}
BE(N_{2}-N,Z_{2}-Z)-BE(N_{1}-N,Z_{1}-Z)\approx a\cdot Z+b\cdot N+c\cdot
Z^{2}+d\cdot N^{2}+e\cdot ZN,  
\label{expansion}
\end{equation}

\noindent where $a,b,c,d,$ and $e$ are constants from the Taylor expansion.
Evaluating Eq. \ref{expansion} within the context of a liquid drop model,
one finds that the second order terms are of the order (1/A), where A is the
mass number, relative to the first order terms. The leading order
parameters, $a$ and $b$ can be interpreted as the differences of the neutron
and proton separation energies for the two compound systems, i.e. $a=$ $%
-\Delta s_{n}$ and $b=-\Delta s_{p}$. Equation \ref{eq:R21_Qgg} can then be
approximated as

\begin{equation}
R_{21}(N,Z)\propto \exp [(-\Delta s_{n}\cdot N-\Delta s_{p}\cdot Z)/T].
\label{eq:R21_Qgg_sep}
\end{equation}

\noindent Eq. \ref{eq:R21_Qgg_sep} confirms the earlier studies which showed
that the symmetry contribution in $\Delta s_{n}$ of the various isotopes
associated with the same element, is approximately linear in the number of
neutrons transferred. Similarly, $\Delta s_{n}$ shows a linear dependence on
the charge transferred due to Coulomb-barrier effects  [10, 11]. Comparison
of Eqs. \ref{eq:alphabeta} and \ref{eq:R21_Qgg_sep} reveals that the
difference in the average separation energies, $\Delta s_{n}/T$ and $\Delta
s_{p}/T$ , plays a corresponding role to the fitting parameters of $\alpha $
and $\beta $. From the binding energy expansion in Eq. \ref{expansion}, one
expects that Eq. \ref{eq:R21_Qgg_sep} becomes less accurate and eventually
breaks down leading to a failure in isoscaling when the range of fragment
masses considered becomes large.

To explore how good is the approximation of using the nucleon separation
energies, we calculate $R_{21}$ obtained with Eq. \ref{eq:R21_Qgg_sep} as
well as using the exact expression of Eq. \ref{eq:R21_Qgg}. We use the two
parent systems that describe the $^{16}O$ + $^{197}Au$ and $^{16}O$ + $%
^{232}Th$ reactions corresponding to $(N_{1},Z_{1})=(126,87)$ and $%
(N_{2},Z_{2})=(150,98)$. The deviations, $R_{21}$(Eq. \ref{eq:R21_Qgg_sep})/$%
R_{21}$(Eq. \ref{eq:R21_Qgg}) are plotted in Figure 3 for Li, Be, B, C, N
and O isotopes as a function of the neutron excess (N-Z). Within each
element, the deviations assume a parabola shape with the minima
located at the neutron rich side of the N=Z line. Over the range of nuclei measured
experimentally, $|N-Z|<3$, the overall deviations are less than 
$\pm 5\%$. However, this comparison suggests that isoscaling will break down
for isotopes with large Z and for nuclei with extreme isospin asymmetry. It
appears likely that first order deviations from the scaling behavior can be
corrected using functions similar to the parabola shown in Figure 3.

{\sc II. Compound nucleus decay}

The scaling behavior for fragments evaporated from an excited compound
nucleus has been discussed in ref. [6]. The measured isotope ratios for $%
^{4}He+^{116}Sn$ and $^{4}He+^{124}Sn$ collisions at $E/A=50MeV$ [12] are
plotted in Figure 4 using the same convention of Figure 1 and 2. At back
angles ($\theta =160^{o}$, left panel), the isotope ratios of different
elements have similar slopes and adjacent elements are separated nearly
equi-distance from each other, typical behavior of isoscaling as evidenced
by the best fit solid and dashed lines. However, at forward angles ($\theta
=12^{o}$) where contributions from pre-equilibrium processes become
significant, isoscaling is not well respected as shown in the right panel.
Not only do different elements have different slopes but also the distances
between adjacent isotones vary greatly [12]. In general, there is a tendency
for the slope to become steeper as the fragment mass is increased,
consistent with the heavier elements emitted at lower temperature. However
for the carbon isotope yields, the trend is actually not monotonic with N,
indicating a clear failing of isoscaling. More detailed discussions on the
forward angle data can be found in Ref. [12] suggesting that the failure of
isoscaling may arise from non-equilibrium emission. 

The origin of isoscaling for evaporation process follows similar derivations
involving the expansion of the binding energies in the Taylor series as
described in previous section, resulting in a formula similar to Eq. (\ref
{eq:R21_Qgg_sep}).

\begin{equation}
R_{21}(N,Z)\propto \exp [(-\Delta s_{n}+\Delta f_{n}^{*})\cdot N-(\Delta
s_{p}+\Delta f_{p}^{*}+\Delta \Phi )\cdot Z)/T].  \label{eq:R21_eva}
\end{equation}

where $f_{n}^{*}$ and $f_{p}^{*}$ are the neutron and proton excited free
energy and $\Phi $ is the electrostatic potential at the surface of a
nucleus. A full derivation of Eq. (\ref{eq:R21_eva}) can be found in Ref.
[6].

{\sc III. Multifragmentation Process}

The scaling phenomenon was first observed in multifragmentation process in
the central $^{124}Sn+^{124}Sn$ and $^{112}Sn+^{112}Sn$ collisions[4] as
demonstrated in Figure 1 and discussed in detail in the introduction
section. To obtain guidance of how the nuclei yield ratios may be
systematized, we examine the dependence of the isotopic yields within the
equilibrium limit of the Grand-Canonical Ensemble [13,14]. In this case
predictions for the observed isotopic yield are governed by both the neutron
and proton chemical potentials, $\mu _{n}$ and $\mu _{p}$ and the
temperature $T_i$, plus the individual binding energies, $BE(N,Z)$, of the
various isotopes [11,12].

\begin{equation}
Y_{i}(N,Z,T_{i})=F_{i}(N,Z,T_{i})\cdot \mathrm{exp}(N\cdot \mu
_{n}/T_{i}+Z\cdot \mu _{p}/T_{i})\cdot \mathrm{exp}(BE(N,Z)/T_{i})
\label{eq:chem}
\end{equation}

The factor $F_{i}(N,Z,T_{i})$ includes information about the secondary decay
from both particle stable and particle unstable states to the final ground
state yields. If the two reactions have the same temperature, ($T_{i}=T)$
the binding energy terms in Eq. (9) cancel out. If one further assumes that
the influence of secondary decay on the yield of a specific isotope is
similar for the two reactions, i.e. $F_{1}(N,Z,T)\approx F_{2}(N,Z,T)$, then
we obtain an equation in the same form as Eq. (1): 
\begin{equation}
R_{21}(N,Z)=C\cdot \mathrm{exp}(N\cdot \Delta \mu _{n}/T+Z\cdot \Delta \mu
_{p}/T)  
\label{eq:GC}
\end{equation}
where $\alpha =\Delta \mu _{n}/T$ and $\beta =\Delta \mu _{p}/T$ reflect the
differences between the neutron and proton chemical potentials for the two
reactions and $C$ is an overall normalization constant. $\Delta \mu _{n}$
and $\Delta \mu _{p}$ correspond to $\Delta s_{n}$ and $\Delta s_{p}$ of Eq. 
\ref{eq:R21_Qgg_sep}. Simulations adopting microcanonical and canonical [8]
multifragmentation models show that Eq. \ref{eq:GC} is respected. Recent SMM
model calculations [15] indicate that $\mu _{n}$ and $s_{n\text{ }}$are
closely related ($\mu _{n}\approx -s_{n}$ $+f_{n}^{*})$ for $0\leq
T\leq 3MeV,$ where the decay configurations are mainly binary, but the
connection between $\mu _{n}$ and $s_{n\text{ }}$ becomes increasingly weak
as the role of multifragment decay configurations becomes important. These
calculations also verify the insensitivities of isoscaling to the effect of
sequential decays [8].

{\sc IV Mixed Systems}

The isoscaling described by Eq. \ref{eq:alphabeta} relies on the emission
mechanisms for the fragments in each reaction being described statistically
with some common effective temperature and that distortions from secondary
decays cancel[4,6,8,16,17]. The exhibition of the systematic trends does not
imply that both reacting systems proceed with the same reaction mechanism.
This point was demonstrated in Ref. [18] where isotopic yields of fragments
produced in central $Au+Au$ multifragmentation process at $E/A=35MeV$ [19]
can be related approximately via isoscaling to those produced in lower
multiplicity evaporation process produced in $Xe+Cu$ reactions at $E/A=30MeV$%
[20]. Isoscaling arises because the temperatures for the two reactions are
nearly the same i.e. $T_{1}\approx T_{2},$ [21] even though the emission
mechanisms in the two reactions differ significantly [19,20]. 

For reactions which differ mainly in temperatures, isoscaling is also
destroyed because the binding energy terms in Eq. \ref{eq:chem} do not
cancel even if the effect of sequential decays can be neglected.

\begin{equation}
R_{21}(N,Z)=C\cdot \mathrm{exp}(N\cdot \alpha ^{\prime }+Z\cdot \beta
^{\prime })\exp (BE/T_{2}-BE/T_{1})  
\label{eq:be}
\end{equation}
where $\alpha ^{\prime }$=$\alpha -k\mu _{n2}$ and $\beta ^{\prime }$=$\beta
-k\mu _{p2}$; $k=1/T_{1}-1/T_{2}$. While the new scaling parameters $\alpha
^{\prime }$ and $\beta ^{\prime }$ are related to $\alpha $ and $\beta $,
they do not have simple physics interpretations. The left panel of Figure 5
shows the $R_{21}$ ratios extracted from isotope yields of $^{124}Sn+^{124}Sn
$ and $^{4}He+^{124}Sn$. Even though the comparison of the two $Sn+Sn$
reactions and the two alpha induced reactions exhibit isoscaling as seen in
Figs. 1 and 4, respectively, there is no observable scaling in these systems
with different temperatures.

Isoscaling could be restored if $R_{21}(N,Z)$ in Eq. \ref{eq:be} is
multiplied by the Boltzmann factor with binding energy and temperatures, $%
\exp (k*BE(N,Z))$. Previous studies suggest that the temperature of the
multifragmentation reaction of $^{124}Sn+^{124}Sn$ collision is about 5 MeV
[22] and the temperature of the evaporation reaction of $^{4}He+^{124}Sn$ is
about 3 MeV [12], we obtain $k\approx 0.12.$ In Figure 4, $R_{21}(N,Z)\exp
(0.12*BE)$ obtained from the same data plotted in the left panel exhibit
very nice systematic behavior shown in the right panel. The restored
isoscaling is clearly demonstrated by the dashed and solid lines which are
the best fits through the data points with $\alpha ^{\prime }=0.939.$

Currently, most temperature measurements depend on selected isotope yields
e.g. the T$_{iso}$(HeLi) depends on the yields of $^{3,4}$He and $^{6,7}$Li
[2,23,24] and T$_{iso}$(CLi) relies on the yields of $^{11,12}$C and $^{6,7}$%
Li [23]. Discrepancies in temperature measurements have been observed
between T$_{iso}$(HeLi) and T$_{iso}$(CLi) [23]. Furthermore, temperatures
derived from excited states (T$_{ex}$) disagree with isotope yield
temperatures (T$_{iso}$) obtained from central collisions at incident energy
greater than 35 MeV [23,24]. Such discrepancies could arise if the light
charged particles with $Z\leq 2$ are emitted early and/or the emitting
sources are not thermalized [20]. With the temperature corrected isoscaling
(Eq. \ref{eq:be}), the internal consistency of the temperature measurements
and the degree of thermalization as a function of incident energy can be
investigated further using all the isotopes measured instead of selected
isotopes.

In summary, we have surveyed many reactions with different reaction
mechanisms. We found that isoscaling occurs if both reactions can be
described by statistical reaction mechanisms and that the temperatures of
both reactions are nearly the same. However, isoscaling does not yield any
information about the reaction mechanisms. In order to draw correct
conclusions from isotopic measurements, it is therefore absolutely essential
to obtain additional experimental information that elucidates the underlying
reaction mechanism. If the temperatures for both reactions are different,
isoscaling can be restored with appropriate temperature corrections. This
work was supported by the National Science Foundation under Grant Nos.
PHY-95-28844 and PHY-96-05140.

\newpage

\textbf{Figure Caption:}

\textbf{Figure 1}Nuclei yield ratios are plotted as
a function of $N$ (top panels) $Z$ (bottom panels) for central $%
^{124}Sn+^{124}Sn$ and $^{112}Sn+^{112}Sn$ collisions at $E/A=50MeV$. The
lines in the upper left panels correspond to best fits of different elements
with one common slope. Similarly, in the bottom left panels, the lines
correspond to fits of the same isotones. In the top right panel, the scaled
isotopic ratio, $S(N)$ (Eq. 2) is constructed using $%
\beta =-0.417$. Similarly, in the bottom right panel, the scaled isotone
ratio, $S(Z)$ defined in Eq. 3, is plotted as a function of $Z$ using 
$\alpha =0.361$.

\textbf{Figure 2}\ Relative isotope ratios for four
systems:\ a.)$^{16}O$ + $^{232}Th$ [9] and $^{16}O$ + $^{197}Pb$ [10] at
incident energy of 137 MeV and $\theta =40^{o}$ (upper left panel). b.) $%
^{14}N$ + $^{100}Mo$ and $^{14}N$ + $^{92}Mo$ at 97 MeV and $\theta =25^{o}$
[9] (upper right panel). c.) $^{22}Ne$ + $^{232}Th$ and $^{22}Ne$ + $^{94}Zr$
at 173 MeV and $\theta =12^{o}$ [9] (lower left panel) and d.) $^{16}O$ + $%
^{232}Th$ and $^{16}O$ + $^{197}Au$ at 315 MeV and $\theta =40^{o}$ [10]
(lower right panel).

\textbf{Figure 3}\ Deviations in approximating $R_{21}$
calculated with Eq. \ref{eq:R21_Qgg_sep} from values calculated with Eq. \ref
{eq:R21_Qgg} plotted as a function of the neutron excess for lithium to
oxygen isotopes for the deeply inelastic reactions of $(N_{1},Z_{1})=(126,87)
$ and $(N_{2},Z_{2})=(150,98)$.

\textbf{Figure 4}\ Relative isotope (Z=3-6) ratios for $%
^{4}He+^{116}Sn$ and $^{4}He+^{124}Sn$ systems emitted at backward 
(left panel) and forward angles (right panel). See Figures 1 and 2 
for symbol conventions.

\textbf{Figure 5}\ Left panel: Disappearance of
isoscaling in reactions with different temperature. Right panel: Isoscaling
is restored if the binding energy terms in the isotope ratio is corrected
for the temperature difference. See Eq. (\ref{eq:be}).

\newpage

\textbf{References:}

1. Isospin Physics in Heavy-Ion Collisions at Intermediate Energies, Eds.
Bao-An Li and W. U. Schroeder, Nova Science Publishers, Inc. (2001).

2. S. Das Gupta, A.Z. Mekjian and M.B. Tsang, Adv. Nucl. Phys. 26 (in press)
and references therein.

3. H. M\"{u}ller and B. D. Serot, Phys. Rev. C 52, 2072 (1995).

4. H.S. Xu, M.B. Tsang, T.X. Liu, X.D. Liu, W.G. Lynch, W.P. Tan and G.
Verde, L. Beaulieu, B. Davin, Y. Larochelle, T. Lefort, R.T. de Souza, R.
Yanez, V.E. Viola, R.J. Charity, L.G. Sobotka; Phys. Rev. Lett. 85, 716
(2000).

5. Bao-An Li et al., Phys. Rev. Lett. 78, 1644 (1997); I. Bombaci, et al.,
Phys. Rep. 242, 165 (1994); J.M. Lattimer and M. Prakash, Ap. J. (in press).

6. M.B. Tsang, W.A. Friedman, C.K. Gelbke, W.G. Lynch, G. Verde, H. Xu,
Phys. Rev. Lett. 86, 5023 (2001).

7. W.P. Tan et al., MSUCL1198 (2001).

8. M.B. Tsang et al., MSUCL1203 (2001).

9. V.V. Volkov, Phys. Rep. 44, 93, (1978) and references therein.

10. C.K. Gelbke et. al., Phys. Rep. 42, 311 (1978) and references therein.

11. J.P. Bondorf, F. Dickmann, D.JH.K. Gross and P.J. Siemens, J.de Phys.
32,\ C6 (1971).

12. J. Brzychczyk et al., Phys. Rev. C47, 1553 (1993).

13. J. Randrup and S.E. Koonin, Nucl. Phys. A 356, 223 (1981);

14. S. Albergo, S. Costa, E. Costanzo, A.Rubbino, Nuovo Cimento A 89, 1
(1985).

15. S.R. Souza, W.P. Tan, R. Donangelo, C.K. Gelbke, W.G. Lynch, M.B. Tsang,
Phys. Rev. C62, 064607 (2000).

16. Y. Murin et al., Europhys. Lett. 34, 337 (1996); Y. Murin et al., Physca
Scripta 56, 137 (1997).

17. O.V. Lozhkin et al., Phy. Rev. C 46, 1996 (1992) and references therein.

18. M.B. Tsang et al., MSUCL1165 (2000).

19. M.J. Huang et al., Phys. Rev. Lett. 78, 1648 (1997).

20. H.F. Xi et al., Phys. Rev. C57, R467 (1998).

21. H. Xi et al., Phys. Lett. B431, 8 (1998).

22. G.J. Kunde, et al., Phys. Lett. B416, 56 (1998).

23. H.F. Xi et al., Phys. Rev. C58, R2636 (1998).

24. V. Serfling et al., Phys. Rev. Lett. 80, 3928 (1998).

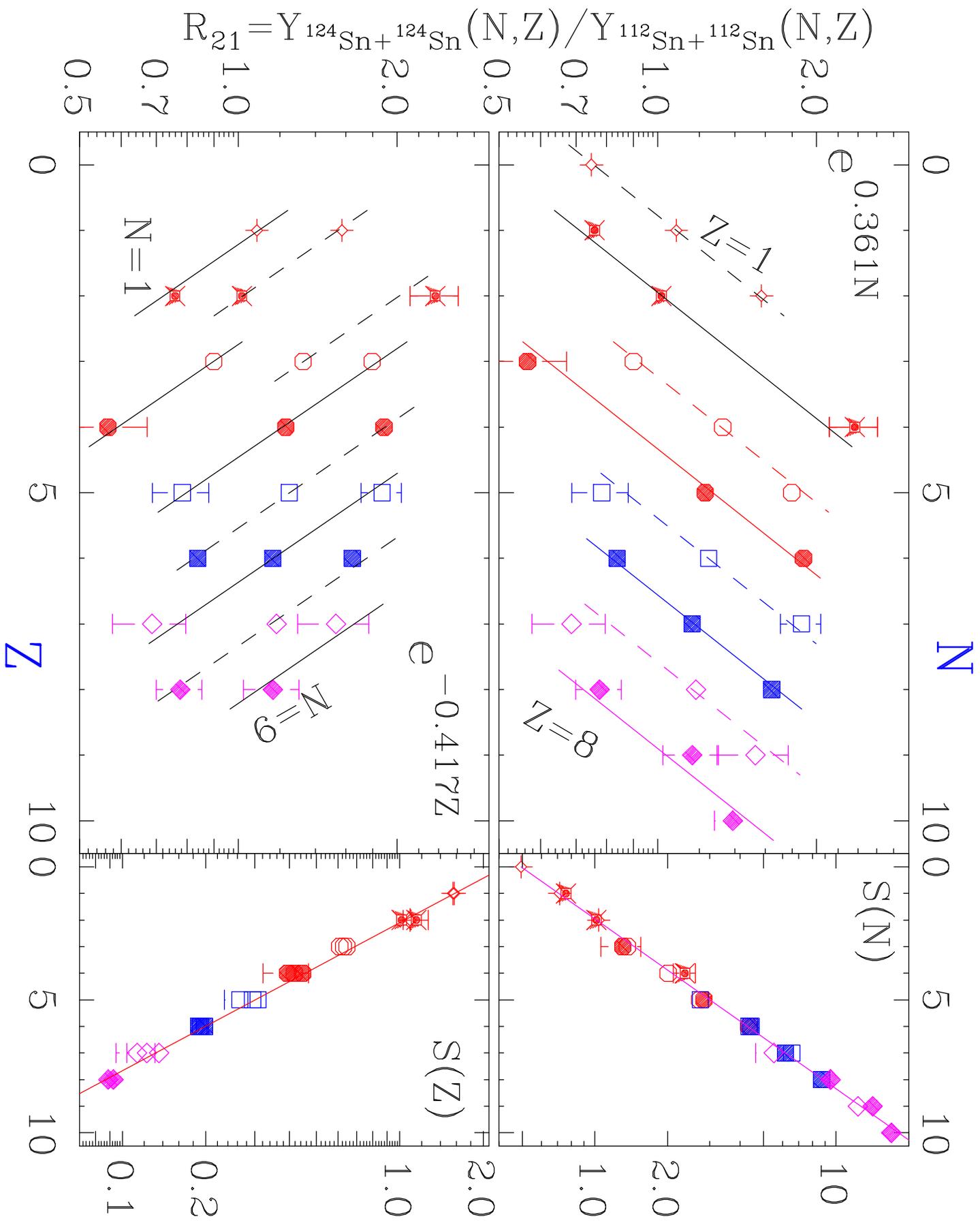
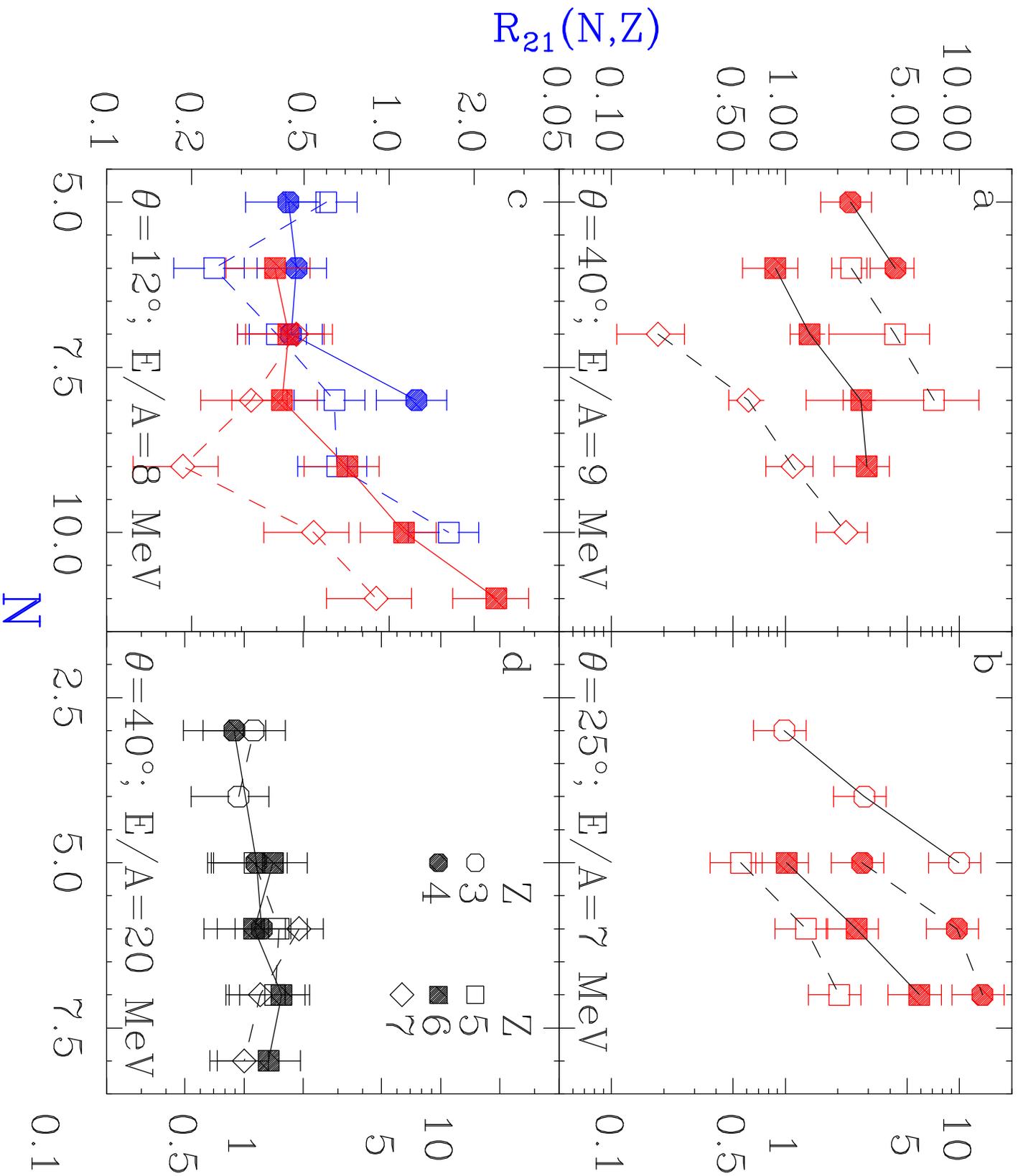
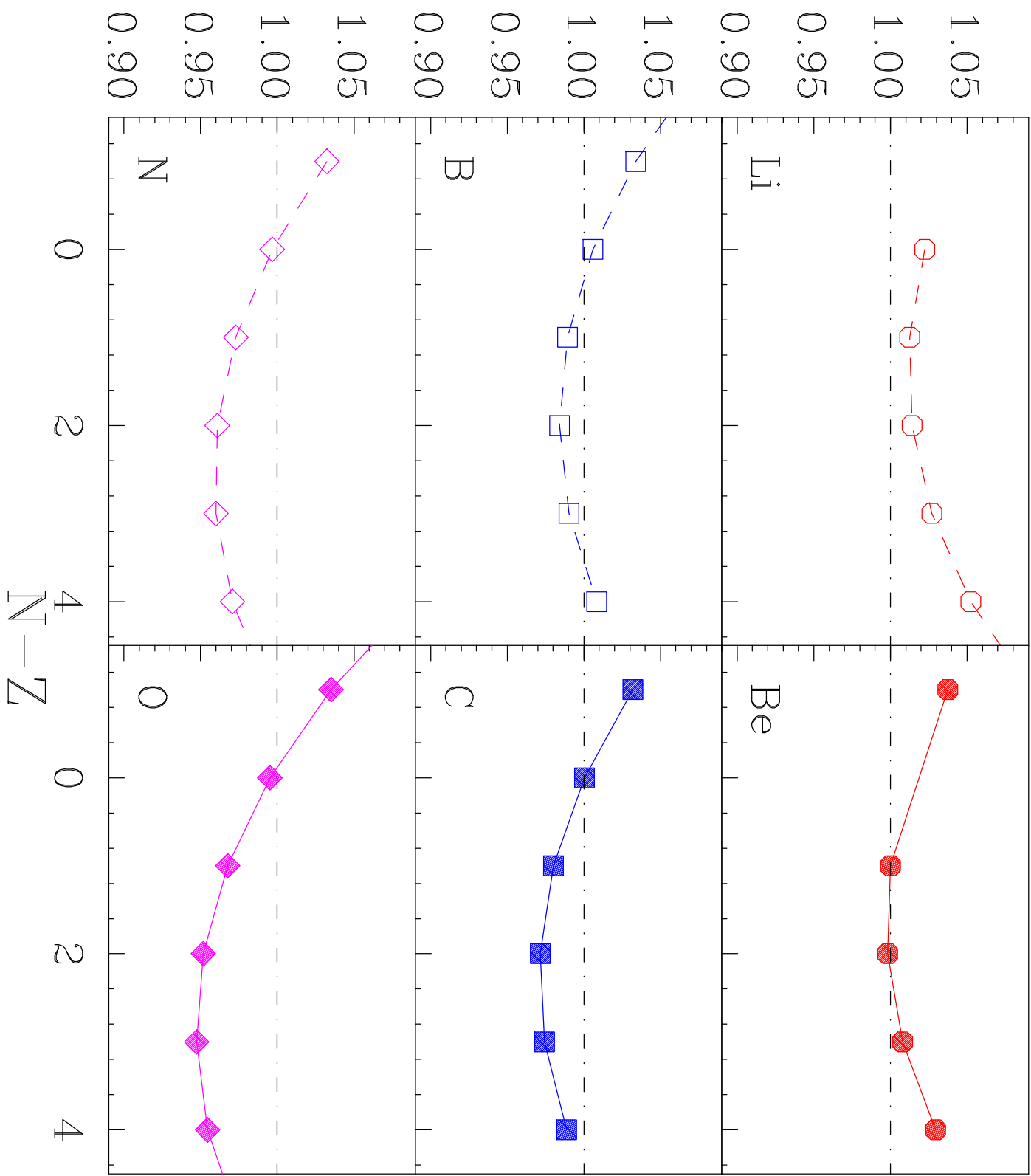
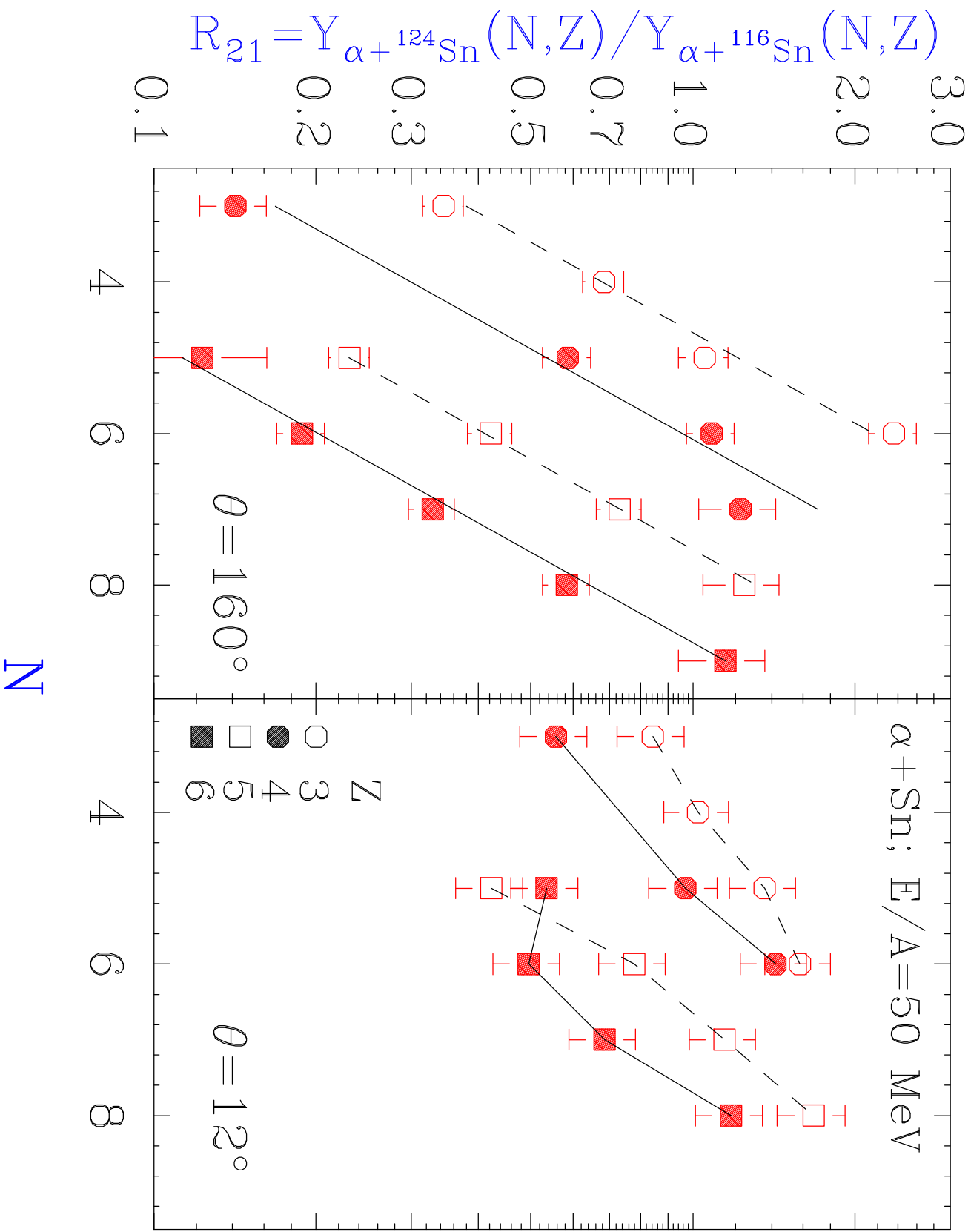
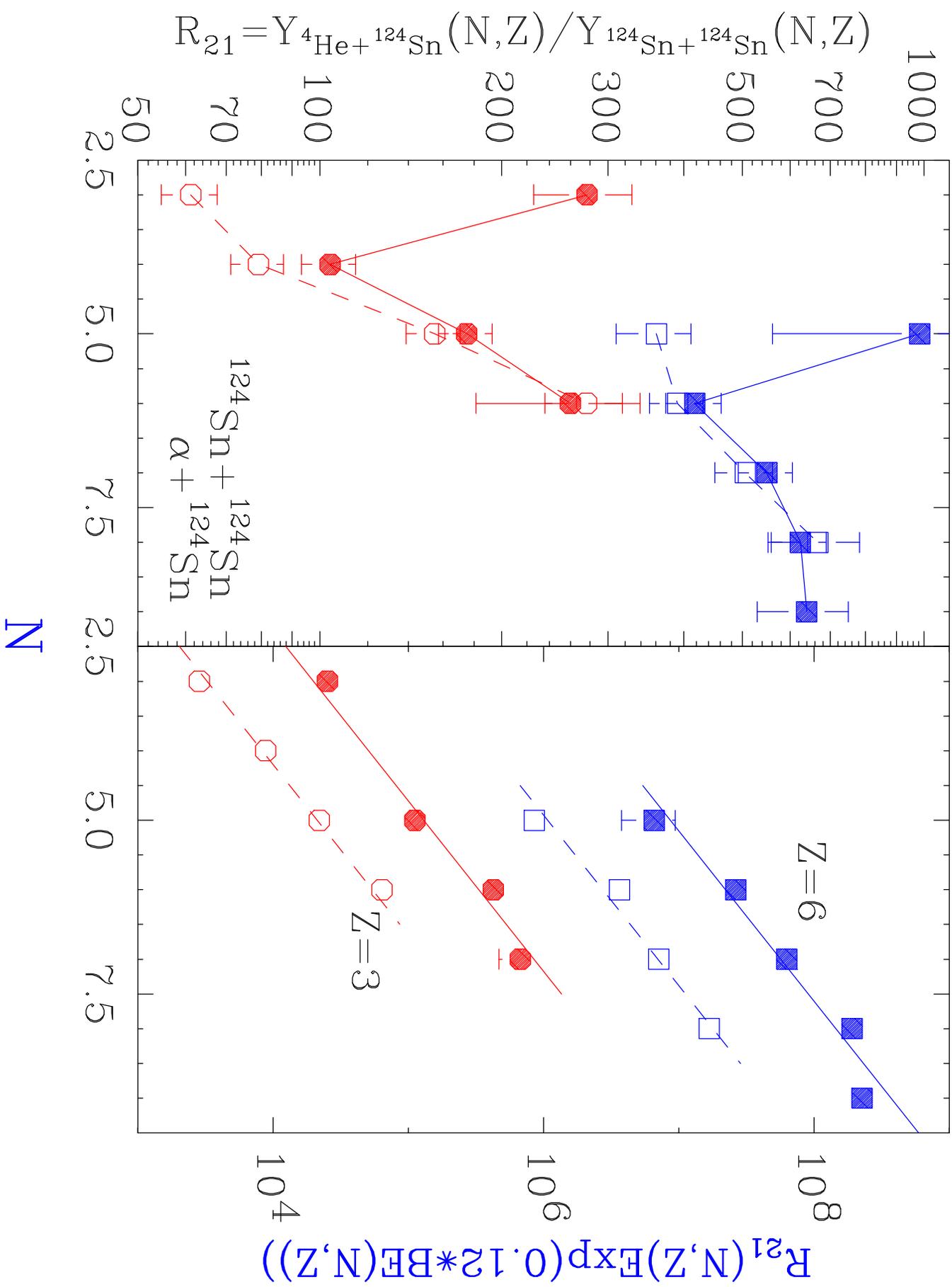
\end{document}